\newcommand{\slot}{\underline{\ \ }\,}
\documentclass[preprint,prd,eqsecnum,aps,tightenlines,12pt]{revtex4}

\begin{document}
\tighten

\def\ie{{\sl i.e.~}}
\def\cf{{\sl cf.~}}
\def\eg{{\sl e.g.~}}
\def\etc{{\it etc.}}
\setlength{\baselineskip}{.3in}

\title{Slowly Rotating Relativistic Stars:\\
VII. Gravitational Radiation from the Quasi-Radial Modes}

\author{{\bf James B.~Hartle}\thanks{hartle@cosmic.physics.ucsb.edu}\\
{\sl Department of Physics,\\
 University of California, Santa Barbara, CA
93106-9530 USA}\\
and\\
{\bf Kip S.~Thorne}\thanks{kip@tapir.caltech.edu}\\
{\sl TAPIR: Theoretical Astrophysics including Relativity, California Institute of Technology,
Pasadena, CA 91125 USA}}

\date{Research completed in 1972; this manuscript completed and put on 
the  arXiv in 2025}

\begin{abstract}
When a relativistic star rotates slowly and rigidly, centrifugal forces flatten it slightly, thereby catalyzing a small admixture of quadrupolar vibration into its radial modes and a damping of the resulting quasi-radial modes.  The damping rate $1/\tau$ of each quasi-radial mode divided by its frequency $\sigma^{(0)}$ is given by  
\[
{1/\tau \over \sigma^{(0)} }= \beta (\sigma^{(0)})^3 \Omega^4 R^8/M \;,
\nonumber
\]
where $\Omega$, $R$ and $M$ are the  angular velocity, radius and mass of the star, and $\beta$ is a dimensionless number that depends on the mode, on relativistic corrections, and on the structure of the star, and is typically of order unity for the fundamental quasi-radial mode in the nonrelativistic limit.  In this paper we develop equations and an algorithm for computing numerically the emitted waves and the resulting damping factor $\beta$.  The rotation is treated to second-order in the angular velocity and the pulsation amplitudes
are assumed small and linearized, but no other approximations are made.

\vskip .3in
\centerline{\noindent\fbox{\parbox{5in}{
{\bf Note by Kip Thorne: }
The research for this paper was completed in 1972, and Jim and I  began writing it in that same year but then got distracted for several decades by other projects.   In 2000, Jim completed a first draft and gave it to me with some public fanfare on the occasion of my 60th Birthday Symposium. I then let it languish until Jim`s 80th Birthday Symposium in 2019, when I gave him a second draft with similar public fanfare.  Jim died on May 17, 2023 --- a huge loss for science and a huge personal loss for me and his many other  friends and admirers.   To my surprise. it appears to me that in the half century since we carried out our analysis, there has been no analysis of this problem by anyone else, so I am making ours available on the Physics arXiv. But I caution readers that Jim and I have not checked the algebraic expressions in Appendix C with the care that we should have, and I no longer have the tools to do so.   To emphasize that caution, I have left in the manuscript just before equation (2.1) a  2019 personal note from me to Jim about two errors that I found and hoped he would check, but did not before his death.  Caveat emptor!}} }
\end{abstract}

\maketitle

\tighten
\vfill\eject
\section{Introduction}
\label{sec: I}

This paper analyzes the gravitational radiation from fully relativistic
stars that are rotating slowly and rigidly, and are pulsating in a mode
(called a quasi-radial mode) that would be radial if the star were not
rotating. These modes radiate only because centrifugal
flattening couples quadrupolar motions into the radial vibrations,
and so
for slow rotation they can be expected to have the longest lifetimes of
all the star's modes.

In the nonrelativistic limit, the centrifugal flattening is described by a stellar ellipticity 
\begin{equation} e \sim \Omega^2 R^3/M 
\end{equation} 
and static quadrupole moment $Q_s \sim M R^2 e$, where $\Omega$, $R$ and $M$ are the star's rotational angular velocity, radius and mass.  Before rotation, the  radial pulsation entails a radial displacement of magnitude $\xi$ and angular frequency $\sigma^{(0)}$, and a pulsation energy $E_{\rm puls} \sim M (\xi \sigma^{(0)} )^2$.  In the rotating star these pulsations induce a time-varying part of the quadrupole moment $Q\sim MR\xi e$ and the emission of axisymmetric gravitational waves (spherical harmonic order $\ell=2$, $m=0$) with power output and pulsation energy loss $dE_{\rm GW}/dt =-dE_{\rm puls}/dt \sim (\partial^3 Q/\partial t^3)^2 \sim (MR\xi e (\sigma^{(0)})^3)^2$.  Correspondingly, the pulsation's energy decays at a rate $2/\tau \sim (dE_{\rm puls}/dt) (E_{\rm puls})^{-1} \sim M R^2 e^2 (\sigma^{(0)})^4$, whence the amplitude decay rate divided by the frequency of pulsation is 
\begin{equation} 
{1/\tau \over \sigma^{(0)}}\sim MR^2 e^2 (\sigma^{(0)})^3 \sim (\sigma^{(0)})^3 \Omega^4 R^8/M \;.
\end{equation}   
This motivates our parametrization of the decay rate in the fully relativistic regime, as given in the abstract: 
 \begin{equation}
{1/\tau \over \sigma^{(0)} }= \beta (\sigma^{(0)})^3 \Omega^4 R^8/M \;,
\end{equation}
where  $\beta$ is a dimensionless number that depends on the mode, on relativistic corrections, and on the structure of the star, and that is typically of order unity for the fundamental quasi-raidal mode in the nonrelativistic limit.

In this paper we  derive an expression which permits the evaluation of a quasi-radial mode's gravitatinal-wave power output
 $dE_{\rm GW}/dt$, and thence  damping rate $1/\tau$ and damping coefficient $\beta$,  as a simple quadrature over a product
of functions
that describe a pulsating, non-rotating star and functions that describe
a rotating, non-pulsating star.

Our results are based on the general analysis of slowly rotating, pulsating
relativistic stars given in Paper VI of this series \cite{HTC73}.
In \S II we take equations derived in Paper VI (particularly Sec.\ H
of Table 3) and cast
them into a form suitable for a calculation of the grafirational-wave power output
from a quasi-radial mode. In \S III we describe how to carry out
such a calculation. 

\section{{\boldmath$\ell=2$} Equations of Motion}
\label{sec: II}

In Paper VI \cite{HTC73}
we constructed the metric, the stress-energy tensor, and the
Ricci tensor for a pulsating and slowly-rotating star. Our analysis was
accurate to second-order in the rigid angular velocity of rotation, 
O$(\epsilon^2)$, and to
first-order in the amplitude of the (adiabatic) pulsations, O$(\zeta)$. 
We also reduced
the $\ell=0$ and $\ell=1$ parts of the Einstein equations to manageable form;
and we derived from them an explicit formula for the rotation-induced
change in the star's pulsation frequency.

In this paper, we turn our attention to the $\ell=2$ parts of the Einstein
equations.  They govern the coupling of the star to its gravitational
waves.   Our analysis and discussion pick up directly where Paper VI left
off, {\it using precisely the same notation}.  Equations in Paper VI will be
denoted ``equation (VI, 4.3a)'' \etc; tables in Paper VI will be denoted
``Table VI-1'', \etc

The final $\ell=2$ equations will be simpler if we use a new variable
$Y(t,r)$ in place of $V(t,r)$.  [Recall that $V$ is the amplitude of the
non-radial fluid displacement $\xi^\theta$; \cf (VI, 3.9).] 
The new variable
$Y$ is defined by (see \cite{Kip2019} for errors Kip found and corrected in this definition)   
\begin{eqnarray}
Y&\equiv& - 8\pi e^{\nu/2} \times \left[{\rm homogeneous},\ \ell=2\ {\rm part\ of
\ Langrangian\ change\ in\ pressure}\right]\nonumber\\
&=& 8\pi e^{\nu/2} \Gamma P \left(r^{-2} e^{-\lambda/2} W^\prime_2 + 6V - K_2 -
\frac{1}{2}\, H_2\right)\ .
\label{twoone}
\end{eqnarray}
(By ``homogeneous part'' we mean those quadrupolar
terms of order $\epsilon^2\zeta$ that 
contain as factors the
variables $H_0, H_2, K_2, N_0, N_2, Q_2, W_0, W_2, V$, instead of
being products of a O$(\zeta)$ spherical vibration function 
with O$(\epsilon^2)$ rotation functions; the latter are the terms
that \emph{drive} the homogeneous functions.)  
In terms of $Y$, $V$ is
given by [invert equation (\ref{twoone})]
\begin{equation}
V=  \frac{1}{6}\, (8\pi \Gamma P)^{-1}\, e^{-\nu/2}\, Y - \frac{1}{6}
\ r^{-2} e^{-\lambda/2} W^\prime_2 + \frac{1}{6}\ K_2 +
\frac{1}{12}\ H_2\ .
\label{twotwo}
\end{equation}

Throughout this paper we shall assume that the star pulsates
sinusoidally in time, so that all time-dependent functions --- $\eta, \mu,
j_1, H_0, H_2, K_2, N_0, N_2, Q_2, W_0, W_2, F, V$, and $Y$ --- have the
form
\begin{eqnarray}
\eta(t, r) &=& \eta (r)\ \exp\left[i\left(\sigma^{(0)} +
\epsilon^2\sigma^{(2)}\right)t\right]\ ,\nonumber\\
N_2(t, r) &=& N_2(r)\ \exp\left[i\left(\sigma^{(0)} +
\epsilon^2\sigma^{(2)}\right)t\right]\ ,\ {\rm \etc}
\label{twothree}
\end{eqnarray}
[\cf equation (VI, 4.5)].

Paper I \cite{Har67}
and \S IV of Paper VI, derived the equations and boundary
conditions for all the functions describing the star except the
$\epsilon^2 \zeta, \ell=2$ functions $H_2, K_2, N_2, Q_2, W_2, V$, and $Y$.
In this section we shall derive equations for these seven remaining functions.

Equation (\ref{twotwo}) expresses $V$ in terms of the other 6 functions.
The functions $H_2$ and $Q_2$ can be eliminated by using some of
the Einstein equations as follows.
To obtain an equation for $H_2$ in terms of $N_2$ and products of functions of 
lower-order than $\epsilon^2 \zeta$, proceed as follows: construct the
$\epsilon^2\zeta \Psi^{2 \ *}_{0 \ \ AB}$  part of the Einstein field equation
$R_{AB}-8\pi (T_{AB} - 1/2\ g_{AB} T)=0$; (see Table VI-3 for the
components of the Ricci tensor); use equation (VI, 4.3a) to eliminate
$j_1^\prime$; use (VI, 4.2a,b) to eliminate $\mu$ and $\eta^\prime$; use
(II, 23a) in \cite{HT68} to eliminate $m_2$; and use (VI, 4.1) to simplify 
terms involving
derivatives of $\nu, \lambda$, and $P$. 

To obtain an equation for $Q_2$ in terms of $N_2, K_2$, and products of 
functions of
lower-order, proceed as follows: construct the $\epsilon^2\zeta {{\Psi^2_0}^*}_A$ 
part of the Einstein equation $R_{tA}=8\pi(T_{tA}- 1/2\ T\ g_{tA})$ (see
Table VI-3 for the components of the Ricci tensor); use the equation
obtained above to express $H_2$ in terms of $N_2$; use (VI, 4.3a) to eliminate
$j^\prime_1$; use (VI, 4.2a,b) to eliminate $\mu$ and $\eta^\prime$; use
(II, 23a) to eliminate $m_2$; and use (VI, 4.1) to simplify terms involving
$\nu, \lambda$, and their derivatives.  

The four remaining functions --- $W_2, K_2, N_2$, and $Y$ --- satisfy 
four coupled differential equations (to be derived below). This coupled system
is of fifth-order:  it involves $K_2^{\prime\prime}, K^\prime_2,
W^\prime_2, N^\prime_2$, and $Y^\prime$.  To facilitate the analysis and
solution of this system, introduce a new variable $Z_2$ defined by
\begin{equation}
 Z_2 \equiv K^\prime_2\ .
\label{twofour}
\end{equation}
Then the system becomes five coupled, first-order differential equations for
the five functions $K_2, Z_2, W_2, N_2, Y$. Regard these 5 functions as a
column vector ${\mathbf W}$:
\begin{equation}
{\mathbf W} = {\bf col} \left[K_2, Z_2, N_2, W_2, Y\right]
\label{twofive}
\end{equation}
and write the coupled system in the form
\begin{equation}
d {\mathbf W}/dr = {\mathbf A} \cdot {\mathbf W} + {\mathbf D}\ ,
\label{twosix}
\end{equation}
where ${\mathbf A}$ is a $5\times5$ matrix and ${\mathbf D}$ is a column
vector of ``driving terms''.

The first term in the coupled system is equation (\ref{twofour}); in the
notation of equation (\ref{twosix}) it says
\begin{equation}
K^\prime_2 = A_{K K} K_2 + A_{K Z} Z_2 + \cdots + D_K\  ,
\label{twoseven}
\end{equation}
in what should be an obvious notation for the components of ${\mathbf
A}$ and ${\mathbf D}$.
Direct comparison with equation (\ref{twofour}) shows that
\begin{equation}
A_{K Z} = 1\ ,\quad A_{K K}=A_{K W} = A_{K N} = A_{K
Y} = D_K = 0\ .
\label{twoeight}
\end{equation}

The remaining components of the matrix ${\mathbf A}$ can be deduced from 
equations~(14) of \cite{TC67}
(which solved our homogeneous equations $d \mathbf W/dr = \mathbf A \cdot
\mathbf W $ in its study of nonradial pulsations of nonrotating stars). 
To deduce $\mathbf A$, first note
that $\{H, K, W\}$ in \cite{TC67} are denoted $\{N_2, K_2, W_2\}$ here,
and that $l$ in \cite{TC67} is 2 here (admixture of quadrupolar vibrations), 
and that
in equation (\ref{twotwo}) above for $V$, $H_2$ can be set equal to $N_2$ for
this homogeneous part of the problem [cf.\ equation (8d) of \cite{TC67},
which we shall denote(\cite{TC67},8d)]. 
Then:  (i) Equation (\cite{TC67},14d), with $V$ substituted by 
(\ref{twotwo}) above takes the form $dW_2/dr = \ldots$, from which we
can read off $A_{WJ}$ for the various $J$.  (ii) Equation (\cite{TC67},14b), 
with $V$ substituted by (\ref{twotwo}) above and $dW_2/dr$ substituted
from (i), gives us $d^2K_2/dr^2 \equiv dZ_2/dr = \ldots$, from which we
can read off $A_{ZJ}$ for the various $J$.  (iii) Equation (\cite{TC67},14a)
with substitutions from equation (\ref{twotwo}), (i) and (ii), gives us
$dN_2/dr = \ldots$, from which we read off the $A_{NJ}$'s.  (iv)
Equation (\cite{TC67},14c), with substitutions from 
equations (\ref{twoone}), (\ref{twotwo}),
(i), and (iii), gives us $dY/dr = \ldots$, from which we read off the
$A_{YJ}$'s.  The results for the components of the matrix $\bf A$ are given
in Appendix B.  

We now turn to the
all-important driving terms, which can be derived beginning with the
Ricci and stress-energy tensors of Paper VI.

The first term in the coupled system is the equation $K'_2 = W_2$, for
which the driving term $D_K$ vanishes [equation\ (\ref{twoeight})].

The second term in the coupled system is an equation for $Z_2^\prime =
K^{\prime\prime}_2$.  To obtain it in explicit form, take the
$\epsilon^2 \zeta\Phi^{2~*}_{0~ AB}$ part of the Einstein equation
$R_{AB}-8\pi(T_{AB}- 1/2\ g_{AB} T)=0$; eliminate $H_2$ and $Q_2$ as
described above; use (VI, 4.3) to
eliminate $j^\prime_1$ and $F$; use the equations of structure for a
rotating star in Paper II \cite{HT68}
to simplify terms involving $\bar\omega, m_2, h_2,
v_2, p^*_2$ and their derivatives; use equation (VI, 4.2) to eliminate
$\mu$ and $\eta^\prime$; and use the equations of structure for a
non-rotating star (eqs.~VI, 4.1) to simplify terms involving derivatives of
$\nu, \lambda$, and $P$.  The result is the $Z_2$ component of equation
(\ref{twosix}), from which one reads off the 
driving term $D_Z$ given in Appendix C.

The third term in the coupled system is an equation for $W^\prime_2$. To
obtain it in explicit form, take the $\epsilon^2 \zeta \Phi^{2~*}_{0~A}$ 
part of
$T^\alpha_{A^; \alpha} =0$; and perform the same substitutions and
simplifications as were used on the $Z^\prime_2$ equation (see above). From
the resulting equation, read off 
the driving term $D_W$ also given in Appendix C.

The fourth term in the coupled system is an equation for $N^\prime_2$. To
obtain it in explicit form, take the $\epsilon^2\zeta\ Y^{2*}_0$ part of the
Einstein equation $G_t^t - 8\pi T_t^t=0$; use the $W_2^\prime$ part of the
coupled system (\ref{twosix}) to eliminate $W^\prime_2$; and perform the
same substitutions and simplifications as were used on the $Z^\prime_2$
equation (see above).  From the resulting equation, read off
$D_N$ as given in Appendix C.

The fifth term in the coupled system is an equation for $Y^\prime$.  To
obtain it in explicit form, take the $\epsilon^2\zeta\ Y^{2*}_0$ part of
$T^\alpha_{r;\alpha} =0$; use the $Z^\prime_2, W^\prime_2$ and
$N^\prime_2$ parts of the coupled system (\ref{twosix}) to eliminate 
$Z^\prime_2, W^\prime_2$ and
$N^\prime_2$; 
and perform the same substitutions and simplifications as
were used on the $Z^\prime_2$ equation (see above). From the resulting
equation, read off $D_Y$ as given in Appendix C.

This large set of algebraic manipulations was carried out with the FORMAC
algebraic computing program on IBM 7094 and IBM 360
computers in the early 1970s.  The results, in
Appendix C, are the nonzero entries in the column vector of driving terms
\[
{\mathbf D} = {\rm\bf col} [DK,\, DZ,\, DN,\, DW,\, DY]\;.
\]

The boundary conditions on the coupled system (\ref{twofive}) will be
discussed in the next section, along with a method for solving the system.

We now have in hand equations to determine {\it all} of the functions which
describe the rotating, pulsating star. 

\section{Gravitational Radiation from a Quasi-Radial Mode}
\label{sec: III}

Equations (\ref{twosix}) for the five remaining unknown $\ell=2$ functions
are to be solved subject to two boundary conditions. The first is that the
solution be suitably regular at the origin. The second is that at large
values of $r$ it contain no incoming gravitational waves. From the
strength of the large $r$ solution, the rate of gravitational radiation can
be calculated. 
In the rest of this section we will show how to impose
these boundary conditions and how to express the rate of gravitational
radiation as a quadrature over the driving terms ${\mathbf D}$ and the
functions which describe a pulsating but non-rotating star.

The key to the method is to construct a Green's function solution to
equation (\ref{twosix}). That is, we write
\begin{equation}
{\mathbf W}(r) = \int^R_0 dr^\prime {\mathbf G} (r, r^\prime)\cdot {\mathbf D}
(r^\prime)\;,
\label{threeone}
\end{equation}
where the $5\times 5$ matrix Green's function ${\mathbf G}$ satisfies
\begin{equation}
\frac{d{\mathbf G}(r, r^\prime)}{dr} - {\mathbf A}(r) \cdot {\mathbf G}(r,
r^\prime) = \delta(r-r^\prime) {\mathbf I}\ ,
\label{threetwo}
\end{equation}
${\mathbf I}$ is the $5\times 5$ unit matrix, and $\cdot$ denotes
matrix multiplication. In writing out
equation (\ref{threeone}), we have taken advantage of the fact that the driving
terms vanish outside the star so the integration need be extended only
to its radius $R$. The boundary conditions on ${\mathbf W}$ can be
enforced by a correct choice of the Green's function.

The boundary conditions are that $\mathbf W(r)$ be 
regular at the origin
and represent an outgoing wave at infinity. To ensure this, the Green's
function must itself have these
properties. To obtain the Green's function, we follow the standard procedure.  Let ${\mathbf
H}(r)$ denote any square matrix of five linearly independent column vector
solutions to the homogeneous tensor equation
\begin{equation}
\frac{d{\mathbf H}(r)}{dr} = {\mathbf A}(r)\cdot {\mathbf H}(r)\ .
\label{threethree}
\end{equation}
The general solution to this equation can be written as a linear
combination of the five independent solutions in ${\mathbf H}$.  That is,
it can be written as ${\mathbf H} \cdot {\mathbf C}$ where ${\mathbf C}$ is a
matrix of constants.

Since the Green's function satisfies the homogeneous equation 
(\ref{threethree}) except at
$r=r^\prime$, it may be written
\begin{eqnarray}
{\mathbf G}(r, r^\prime) &=& {\mathbf H}(r)\cdot {\mathbf
V}(r^\prime)\qquad r<r^\prime\ ,\nonumber\\
&=& {\mathbf H}(r)\cdot {\mathbf U}(r^\prime)\qquad r>r^\prime\ .
\label{threefour}
\end{eqnarray}
for some matrices of coefficients ${\mathbf U}(r^\prime)$ and ${\mathbf
V}(r^\prime)$ to be determined by the boundary conditions.
In order that ${\mathbf G}$ be regular at the origin, it must be made up of
only regular solutions of the homogeneous equation for $ r<r^\prime$. If we
denote by ${\mathbf P}_0$ the projection onto the regular solutions (so
that ${\mathbf H}(r)\cdot {\mathbf P}_0$ has only regular solutions) then
the condition on ${\mathbf V}(r^\prime)$ is:
\begin{equation}
{\mathbf P}_0 \cdot {\mathbf V}(r^\prime) = {\mathbf V}(r^\prime)\ .
\label{threefive}
\end{equation}
If ${\mathbf P}_\infty$ denotes the projection onto solutions of the
homogeneous equation that have no incoming waves at infinity then the
boundary condition that the Green's function have no incoming waves
may be expressed as
\begin{equation}
{\mathbf P}_\infty \cdot {\mathbf U}(r^\prime) = {\mathbf U}(r^\prime)\ .
\label{threesix}
\end{equation}
The final condition which, together with eqs.~(\ref{threefive}) and
(\ref{threesix}), determines ${\mathbf U}$ and ${\mathbf V}$ is found by
integrating equation (\ref{threetwo}) across $r=r^\prime$. One finds
\begin{equation}
{\mathbf H}(r)\cdot {\mathbf U}(r) - {\mathbf H}(r)\cdot {\mathbf V}(r) =
{\mathbf I}\ ,
\label{threeseven}
\end{equation}
or equivalently.
\begin{equation}
{\mathbf U}(r) - {\mathbf V}(r)  = {\mathbf H}^{-1}(r)\ .
\label{threeeight}
\end{equation}

There can be no radiation of a star without a driving oscillation.  Thus,
there can be no solution of the homogeneous equation which is regular at
the origin and has no incoming waves. Mathematically this physical
fact has the expression
\begin{equation}
{\mathbf P}_0 \cdot {\mathbf P}_\infty = {\mathbf P}_\infty \cdot {\mathbf
P}_0 = {\mathbf 0}\ .
\label{threenine}
\end{equation}
Multiplying equation (\ref{threeeight}) by ${\mathbf P}_0$ and using
equation (\ref{threefive}), we can solve for ${\mathbf V}(r)$. Using ${\mathbf
P}_\infty$, ${\mathbf U}$ can be found in a similar fashion;
and then with the aid of equation (\ref{threenine}) we can
simplify these results to $\mathbf V(r) = - {\mathbf P}_0\cdot \mathbf H^{-1}(r)$
and $\mathbf U(r) =  {\mathbf P}_\infty\cdot \mathbf H^{-1}(r)$.  
Inserting these into equation (\ref{threefour}), we find 
for the Green's function: 
\begin{eqnarray}
{\mathbf G}(r,r^\prime) &=& - {\mathbf H}(r)\cdot {\mathbf P}_0 \cdot
{\mathbf H}^{-1}(r^\prime) \qquad r<r^\prime\nonumber\\
{\mathbf G}(r,r^\prime) &=& {\mathbf H}(r)\cdot {\mathbf P}_\infty \cdot
{\mathbf H}^{-1}(r^\prime)\qquad r>r^\prime\ .
\label{threeten}
\end{eqnarray}
It then follows immediately from equation (\ref{threeone}) that for $r>R$
\begin{equation}
{\mathbf W}(r) = {\mathbf H}(r)\cdot {\mathbf P}_\infty \cdot \int^R_0
\ dr^\prime\ {\mathbf H}^{-1}(r^\prime) \cdot {\mathbf D}(r^\prime)\ .
\label{threeeleven}
\end{equation}
The asymptotic behavior of ${\mathbf W}(r)$ can thus be determined from the
asymptotic behavior of the homogeneous solution ${\mathbf H}$ and the
quadrature of ${\mathbf D}$ with $\mathbf H^{-1}$. 

The homogeneous solutions ${\mathbf H}$ have already been extensively
studied.  This is because, in the absence of any rotational driving
terms, the vector equation (\ref{twosix}), which gives
rise to the tensor equation (\ref{threethree}), describes the
non-radial $\ell=2$ oscillations of a non-rotating spherical star.  This
problem has been considered in detail by Thorne and co-workers 
\cite{TC67,TC67a,PT69,Tho69,TI73}.
In this paper we have chosen the gauge so that it coincides with that used
in this series of papers (aside from our renormalization of the spherical
harmonics; see Table 1 of VI). The results derived there may then be taken over directly
here for the homogeneous vector and tensor solutions. There are two linearly-independent
vector solutions which are regular at the origin. They may be defined by their
small $r$ behavior (equation (15a) of \cite{TC67a}):
\begin{mathletters}
\label{threetwelve}
\begin{equation}
\mathbf W \equiv {\bf col}(K_2,Z_2,N_2,W_2,Y) = {\bf col} \left[r^2, 2r, r^2, 0, 0\right]
\label{threetwelve a}
\end{equation}
and
\begin{equation}
\mathbf W = {\bf col}\ \left[0, 0, 0, r^3, -\frac{1}{2}\, r^2\right]\ .
\label{threetwelve b}
\end{equation}
\end{mathletters}

There are two linearly independent solutions which have no incoming
waves at infinity. They may be taken to be the ``outgoing'' and
``standing wave'' solutions in equations (8) and (9)
of \cite{PT69}. 
(After this research was completed, Ipser and Thorne \cite{TI73} discovered
that the ``standing wave'' solution is spurious in a sense that they describe.
In our analysis, as in all earlier analyses \cite{TC67,TC67a,PT69,Tho69},
it is removed by the boundary conditions we impose.)  
Outside the star,
the two equations for the motion of the fluid --- for $Y$ and $W_2$ ---
vanish.  The asymptotic behavior is therefore determined by the behavior of
$[K_2, Z_2, N_2]$. For the outgoing and (spurious) standing wave solutions
respectively, these have the large $r$ behaviors 
\begin{mathletters}
\begin{equation}
\mathbf W = \exp\left\{i\sigma^{(0)}\left[r+2M{\rm \ell n}(r-2M)\right]\right\}\times
{\rm\bf col} [1,\, i\omega r,\, i\omega r , \slot , \slot ]\;,
\label{threethirteen a}
\end{equation}
\begin{equation}
\mathbf W = r^{-6}\ {\rm\bf col} \left[1,\, \left(2\sigma^{(0)}r\right)^2,
\, 2i\sigma^{(0)} r^2, \slot , \slot \right]\ .
\label{threethirteen b}
\end{equation}
\end{mathletters}
Asymptotically the ingoing wave solution is the complex conjugate of
(\ref{threethirteen a}). (The large $r$ divergences of these equations are
an artifact of the Regge-Wheeler gauge as explained in \cite{PT69}.)

There are three boundary conditions necessary to restrict to the two
possible solutions regular at the origin. There is one boundary condition
restricting to no incoming waves. There is one overall normalization for a
total of five conditions completely determining a solution.

If we take the outgoing wave, (spurious) standing wave, incoming wave, 
and two origin-regular 
solutions in that order, as the column vectors composing the matrix
${\mathbf H}(r)$, then the matrices ${\mathbf P}_0$ and ${\mathbf P}_\infty$
are
\begin{mathletters}
\begin{eqnarray}
{\mathbf P}_0 &=& {\rm\bf diag}\ (0, 0, 0, 1, 1)\;,\\
\label{threefourteen a}
{\mathbf P}_\infty &=& {\rm\bf diag}\ (1, 1, 0, 0, 0)\;. 
\label{threefourteen b}
\end{eqnarray}
\end{mathletters}

Calculation of the power radiated in gravitational waves and then the quasi-radial mode's damping rate is now a
straightforward quadrature using the formulae presented here:
\begin{itemize}
\item Solve for the structure of the slowly-rotating, non-pulsating star
and for a radial mode of the non-rotating, pulsating star.

\item Solve the homogeneous equations $d \mathbf W/dr = \mathbf A \cdot \mathbf W$
for the $\ell=2$, non-radial
perturbations of the non-rotating star.  Use these to construct the matrix
${\mathbf H}(r)$ as described above, and thence the Green's function $\mathbf
G(r,r')$ [equation (\ref{threeten})].

\item Integrate (\ref{threeone}) to find  $\mathbf W$, i.e. the $0(\epsilon^2\zeta)$
description of the pulsating, slowly-rotating star.  Focus on
the amplitude $C^{(O)}$ of outgoing radiation as the coefficient of $K_2(r)$ (the first component of $\mathbf W$)
at large $r$:
\begin{equation}
K_2(r)\to C^{(O)}\, \exp\left\{i\sigma^{(0)}[r+2M{\rm \ell n}(r-2M)]\right\}
\label{threefifteen}
\end{equation}

\item Use this coefficient $C^{(O)}$ to evaluate the angular distribution
of radiated power by equations (28) and (16a) of \cite{PT69} (with $\ell=2$, $m=0$
and with $[(2\ell+1)/4\pi]^{1/2} = (5/4\pi)^{1/2}$ removed from
$\Upsilon^2_{0\,(A)(B)}$ because of our renormalization of the
Regge-Wheeler spherical harmonics; see Table 1 of Paper VI).  
The result is
\begin{equation}
\frac{dE_{GW}}{d\Omega dt} = \frac{9\sin^4\theta}{128\pi} \, \left|C^{(O)}\right|^2\,.
\label{threesixteen}
\end{equation}

\item
Integrate this over the sphere (or equally well from equation (30) of \cite{PT69}),
obtain the power radiated:
\begin{equation}
\frac{dE_{GW}}{dt} = - \frac{dE_{\rm puls}}{dt} = \frac{3}{20}  \, \left|C^{(O)}\right|^2\;.
\label{threeseventeen}
\end{equation}

\item
Compute the star`s pulsation energy using the nonrotating-star formula  $E_{\rm puls} = P^* + K^*$ where $P^*$ is potential energy and $K^*$ is kinetic energy, given by Eqs. (B.26) and (B.27) of \cite{HTWW} (but 
with errors corrected:  the constant multiplicative factors $e^{\nu_0 (0)/2}$ must be deleted).  Then evaluate the energy damping rate (twice the amplitude damping rate)
\begin{equation}
{2\over \tau} = \frac{dE_{\rm puls}/dt}{E_{\rm puls}} \;.
\end{equation}

\item
Compare with equation (1.3) to deduce the star's dimensionless damping coefficient $\beta$.

\end{itemize}

\acknowledgments

We thank the Computer Centers at UCSB and Caltech for help and accommodation
of this difficult problem {\it c}. 1968--1972. The work of JH was supported in
part by an Alfred P.~Sloan Research Fellowship.  The research was supported
in part by the NSF, under grants GP-28027 and GP-27304 in the early 1970s, and 
grants PHY95-07065 and PHY-0601459 during JH's completion of the first draft of this paper.

\newpage
\centerline{\bf Appendix A.  Translation Table}
\vskip .10in
This table translates between the notation of Paper VI and the FORTRAN
notation in which the components of $\mathbf A$ (Appendix B) and the
$\ell=2$ driving terms (Appendix C) are given. ({\tt \#E} is $e$
--- the base of the natural logarithm.) 
\vskip .26 in
{\tt
\noindent
Non-rotating, non-pulsating,  0(1)
\vskip .10in
\begin{tabular}{cccccccccccccccccccccccccccccccccccc}
R&&&&&{$r$}&&&&&&&&&&P&&&&&$8\pi P$\\
S&&&&&$\sigma^{(0)}$&&&&&&&&&&EAP&&&&&$8\pi(E+P)$\\
L&&&&&$\lambda$&&&&&&&&&&GES&&&&&$\gamma$\\
NU&&&&&$\nu$&&&&&&&&&&B&&&&&$e^\lambda$\\
E&&&&&$8\pi E$&&&&&&&&&&C&&&&&$e^{-\lambda}$\\
&&&&&&&&&&&&&&&V&&&&&$1/\nu^\prime$ &this V must not be confused with equation (2.2)\\
\end{tabular}
\vskip .26 in
\noindent
Rotating, non-pulsating
\vskip .10in
\begin{tabular}{ccccccccccccccccccccccc}
$0(\epsilon)$&&J1&&&&$j_1$&&&&&&$O(\epsilon^2)$&&H2&&&&&$h_2$\\
                &&OM&&&&$\Omega$&&&&&&               &&V2&&&&&$v_2$\\
                &&MB&&&&$\bar\omega$&&&&&&           &&&&&&&\\
                &&MBP&&&&$\bar\omega^\prime$&&&&&&   &&&&&&&
\end{tabular}
\vskip .26 in
\noindent
Non-rotating, pulsating, $0(\zeta)$
\vskip .10in
\begin{tabular}{cccccccccccccccccccccc}
U&&&&&$U$&&&&&&&&&&GA&&&&&$\Gamma$\\
UP&&&&&$U^\prime$&&&&&&&&&&GAP&&&&&$\Gamma^\prime$\\
ET&&&&&$\eta$&&&&&&&&&&&&&&&
\end{tabular}
}
\newpage

{\hsize=7truein
\hoffset=-.5truein
\footnotesize

\setlength{\textwidth}{6in}
\setlength{\oddsidemargin}{.0in}
\setlength{\evensidemargin}{.5in}


\centerline{\bf \normalsize Appendix B.  The Matrix $\mathbf A$}
\vskip .5in

\small

\begin{verbatim}

AKK=
0

AKZ=
1

AKN=
0

AKW=
0

AKY=
0


AZK=
#E**L*(4/R**2 - S**2/#E**NU)

AZZ=
(-2 + #E**L*(-2 + E*R**2 - P*R**2))/(2*R)

AZN=
(12*#E**NU*GA*P*(-1 + R**2) - 3*#E**(L + NU)*EAP*(EAP + 
GA*P)*R**2*(-1 + R**2) + 2*EAP*R**2*S**2 - 
#E**L*EAP**2*R**4*S**2)/(6*#E**NU*GA*P*R**2*(-1 + R**2) + 
EAP*R**4*S**2)

AZW=
(#E**(L/2)*EAP**2*(-1 + #E**L*(1 + P*R**2))*(6*#E**NU*(GA - 
GES)*P*(-1 + R**2) + (EAP - 
GES*P)*R**2*S**2))/(2*GES*P*R**3*(6*#E**NU*GA*P*(-1 + R**2) + 
EAP*R**2*S**2))

AZY=
-((#E**(L - NU/2)*EAP*(EAP - 
GA*P)*R**2*S**2)/(GA*P*(6*#E**NU*GA*P*(-1 + R**2) + EAP*R**2*S**2)))


ANK=
(#E**L*(2 - (R**2*S**2)/#E**NU))/R

ANZ=
(3 - #E**L*(1 + P*R**2))/2

ANN=
(24*#E**NU*GA*P*(-1 + R**2) + 6*#E**(L + NU)*GA*P*(-1 + R**2)*(-8 - 
E*R**2 + P*R**2) + 4*EAP*R**2*S**2 - #E**L*EAP*R**2*(8 + E*R**2 + 
EAP*R**2 - P*R**2)*S**2)/(2*R*(6*#E**NU*GA*P*(-1 + R**2) + 
EAP*R**2*S**2))

ANW=
-(#E**(L/2)*EAP**2*(-1 + #E**L*(1 + 
P*R**2))*S**2)/(2*(6*#E**NU*GA*P*(-1 + R**2) + EAP*R**2*S**2))

ANY=
(#E**(L - NU/2)*EAP*R**3*S**2)/(6*#E**NU*GA*P*(-1 + R**2) + 
EAP*R**2*S**2)

AWK=
#E**(L/2)*R**2

AWZ=
0

AWN=
(#E**(L/2)*R**2*(6*#E**NU*(EAP*R**2 + GA*P*(-1 + R**2)) + 
EAP*R**2*S**2))/(2*(6*#E**NU*GA*P*(-1 + R**2) + EAP*R**2*S**2))

AWW=
(3*#E**NU*EAP*R*(-1 + #E**L*(1 + P*R**2)))/(6*#E**NU*GA*P*(-1 + 
R**2) + EAP*R**2*S**2)

AWY=
-((#E**((L - NU)/2)*R**2*(6*#E**NU*GA*P - 
EAP*R**2*S**2))/(GA*P*(6*#E**NU*GA*P*(-1 + R**2) + EAP*R**2*S**2)))


AYK=
(EAP*(-#E**NU + #E**(L + NU)*(-1 + P*R**2) + 
#E**L*R**2*S**2))/(2*#E**(NU/2)*R)

AYZ=
(#E**(NU/2)*EAP*(1 + #E**L*(1 + P*R**2)))/4

AYN=
(#E**(NU/2)*EAP*(-6*#E**NU*(EAP + GA*P*(-1 + R**2)) + 6*#E**(L 
+ NU)*(EAP + EAP*P*R**2 + GA*P*(-1 + R**2)*(5 + E*R**2 - 4*P*R**2)) - 
EAP*R**2*S**2 + #E**L*EAP*R**2*(5 + E*R**2 + EAP*R**2 - 
4*P*R**2)*S**2))/(4*R*(6*#E**NU*GA*P*(-1 + R**2) + EAP*R**2*S**2))

AYW=
(#E**((-L - NU)/2)*EAP*(6*#E**(2*NU)*(EAP + 11*GA*P*(-1 + R**2)) + 
6*#E**(2*(L + NU))*(1 + P*R**2)*(EAP + EAP*P*R**2 + GA*P*(-1 + R**2)*(3 - 
2*E*R**2 + P*R**2)) - 12*#E**(L + 2*NU)*(EAP + EAP*P*R**2 + GA*P*(-1 
+ R**2)*(7 - 2*E*R**2 + 3*P*R**2)) + 11*#E**NU*EAP*R**2*S**2 + 
#E**(2*L + NU)*EAP*R**2*(1 + P*R**2)*(3 - 2*E*R**2 + EAP*R**2 + 
P*R**2)*S**2 - #E**(L + NU)*R**2*(EAP**2*R**2 + 24*GA*P*(-1 + R**2) + 
EAP*(14 - 4*E*R**2 + 6*P*R**2))*S**2 - 
4*#E**L*EAP*R**4*S**4))/(4*R**4*(6*#E**NU*GA*P*(-1 + R**2) + 
EAP*R**2*S**2))

AYY=
-(EAP*(-6*#E**NU + 6*#E**(L + NU)*(1 + P*R**2) + 
#E**L*EAP*R**4*S**2))/(2*R*(6*#E**NU*GA*P*(-1 + R**2) + 
EAP*R**2*S**2))



\end{verbatim}

\vfill\eject


\centerline{\normalsize \bf Appendix C.  The Driving Terms}
\vskip .3in

\small 

\begin{verbatim} 
DK= 0

\end{verbatim}
\vskip .3in
 
\begin{verbatim}
DZ= -(8/3)*J1*(MB-OM)*#E**(-NU)*(1-(1/2)*EAP*R**2*#E**L)
    +U*H2*#E**(2*L+NU/2)*(EAP**2*((1-3*#E**L)*V/R**2+2*(1-#E**L)*V**2/R**3-P*#E**L*V*
  (1+2*V/R)+2*V*(1-#E**(-L))/(GES*P*R**4)+(7/2)*(1-#E**(-L))/(GES*P*R**3)+(5/2-2*
  EAP/P)/(GES*R)-5/(2*R)+#E**(-L)*GESP/(GES**2*P*R**2))+EAP**3*(V*#E**L-V/(GES*P*R**2)-
  (1/2)*(1-#E**(-L))/(GES*P**2*R**3)+(1+(1-#E**(-L))/(P*R**2))/(GES**2*P*R))+EAP*(-V*(6*
  #E**(-L)-4-2*#E**L)/R**4+P/(2*R)+4*V**2*#E**(-L)*(#E**L-1)**2/R**5+2*V*P*(1+
  #E**L)/R**2-4*V**2*P*(1-#E**L)/R**3+(19/2-(11/2)*#E**(-L))/R**3+3*GA*P*V/R**2)+GA*P*
  (E/R-(1-#E**(-L))*(6*V/R+1)/R**3))
    +U*MB**2*#E**(-NU/2)*((1/3)*EAP*V*#E**(3*L)*(-2*EAP+E**2*R**2)+(2/3)*EAP*V**2*
  #E**(2*L)*(EAP*(1-#E**L)-E*P*R**2*#E**L)/R-(10/3)*EAP*R*#E**(2*L)*(EAP+(7/20)*(E+2*
  P)*P*R**2)+(3/2)*EAP**2*R*#E**L*(1+(4/9)*EAP*R**2*#E**L)+EAP*#E**L*((#E**(2*L)/3+2*
  #E**L/3-1)*V/R**2+(2/3)*V**2*(#E**L-1)**2/R**3+(2/3)*P*V*#E**(L)*(#E**L+1)-(4/3)*P*
  V**2*#E**(L)*(1-#E**L)/R+(11/3)*P*R*(1+(5/22)*#E**L)+(2/3)*R*S**2*#E**(-NU)*((GA-1)*
  P-E)/(GA*P)-(5/2*#E**(-L)-2*#E**L+27/2)/R)+EAP**2*((-7/6+#E**(2*L)/6+#E**L)/(GA*P*R)+
  R*#E**L*(1+#E**L*(1+P*R**2/2))/(3*GA)-V*#E**L*(1-#E**L)/(3*GES*P*R**2)+(-7/3+(5/6)*
  #E**(2*L)+#E**L/2)/(GES*P*R)-R**3*#E**(2*L)*(E+(1-GA/3)*P)/(2*GES)-GA*R*#E**L*(1-
  #E**L)/(6*GES)+GESP*(1-#E**L*P*R**2)/(3*GES**2*P)+#E**L*R*(4-#E**L*(1/6+P*R**2))/GES-
  V*E*#E**(2*L)/(3*GES*P))+EAP**3*((2/3)*R*#E**L*(1-#E**L/2)/(GES**2*P)-(1-#E**L)/(3*
  GES**2*P**2*R)-R**3*#E**(2*L)/(3*GES**2)+(2/3)*R*#E**L*(1-3*#E**L/2)/(GES*P)+(1-
  #E**L)/(6*GES*P**2*R))+GA*P*(V*E*#E**(2*L)+(P**2-E**2)*R**3*#E**(2*L)/3+(2/3)*EAP*R*
  #E**L*(#E**L-4)+V*#E**L*(1-#E**L)/R**2+#E**L/(3*R)))
    +U*MBP**2*#E**(L-NU/2)*(EAP*(V*((1+#E**L)-E*R**2*#E**L)/6+R*#E**L*P*V**2/3+(#E**L-
  1)*V**2/(3*R)-R*(35*#E**(-L)/12+4*(1+P*R**2/16)/3))+EAP**2*(R**3/3+V/(6*GES*P)+R*
  (#E**(-L)-1/2-P*R**2)/(6*GES*P))+GA*P*(-R**3*(E+P/2)/6-V/2+R*(#E**(-L)+1/3)/4))
    +U*MB*MBP*#E**(-NU/2)*EAP*((8/3)*(E+7*P/4)*R**2*#E**L-2-8*#E**L+R**2*(-4*GA*P/3+
  2*EAP/GES)*#E**L+4*EAP*(#E**L-3/2)/(3*GES*P))
    +U*V2*#E**(2*L+NU/2)*(EAP*(-4*V*E*#E**L/R**2-8*V**2*(1-#E**L)/R**5+4*V*(1+
  #E**L)/R**4+8*P*V**2*#E**L/R**3-6/R**3)+EAP**2*(4*V/(GES*P*R**4)+2/(GES*P*R**3))-12*
  GA*P*V/R**4)
    +UP*H2*#E**(L+NU/2)*(EAP*(V*#E**L*(E+(1-GA)*P)/R+2*V*(1-#E**L)/R**3-3/R**2)-
  EAP**2/(GES*P*R**2)-2*GA*P*V*(1-#E**L)/R**3+GA*(E-P)/R**2)
    +UP*MB**2*#E**(L-NU/2)*(EAP*(#E**L*R*E*V/3+R**2*(E+8*P/3)+V*(1-#E**L)/(3*R)-7*
  #E**(-L)-1-GA*P*R*#E**L*V/3+GA*(1-2*P*R**2)/3+2*GAP*R*#E**(-L)/(3*GA))+EAP**2*(-
  (1-#E**(-L))/(3*P)+(1-2*#E**(-L))/(3*GES*P)+2*R**2/(3*GES))+GA*P*((2/3)*GA*P/EAP+R*
  V*#E**L*P/3+V*(#E**L-1)/(3*R)-P*R**2/3+3*#E**(-L)+4/3))
    +UP*MBP**2/6*#E**(-NU/2)*(R*#E**L*V*((GA-1)*P-E)+R**2*(E+(GA+1)*P))
    +4*UP*MB*MBP*R*#E**(-NU/2)*((GA/3-1)*P-E)
    +4*UP*V2*V*#E**(2*L+NU/2)*((GA-1)*P-E)/R**3
\end{verbatim}
\vskip .5in
\begin{verbatim}
DN= +ET*H2*(-EAP*#E**L*V/2+EAP/2*V**2*R**(-1)*#E**L*(#E**L-1)+EAP*P*R*#E**(2*
  L)/2*V**2+(1-#E**(2*L))/R**3*V**2+2*(#E**L-1)*V**2*R**(-3)-P*V*#E**L-P*#E**(2*L)*
  V**2*R**(-1)+P*#E**L*V**2/R+R**(-1))
    +ET*MB**2*(-EAP*R**2*#E**(L-NU)*V/6+EAP/6*R*V**2*#E**(L-NU)*(#E**L-1)+EAP/6*V**2*
  P*R**3*#E**(2*L-NU)-#E**(-NU)/6*V**2/R*(1-#E**L)**2-1/6*P*R*V**2*#E**(-NU)*(#E**(2*
  L)*(2+P*R**2)-2*#E**L)+1/6*R*#E**(-NU))
    +1/12*ET*MBP**2*R**2*V*#E**(-NU)*(1+V/R-#E**L*V/R*(1+P*R**2))
    -2/3*ET*MBP*(MB-OM)*R**2*#E**(-NU)
    +2*ET*V2*#E**L*V/R*(1/R+R**(-2)*V*(1-#E**L)-P*V*#E**L)
    -8/3*U*MB*OM*EAP*#E**(L-NU/2)
    +2/3*U*MB*MBP*(EAP)*R*#E**(-NU/2)*(4+#E**L-2*GA*P*R**2*#E**L)-2/3*U*MBP*OM*
  EAP*R*#E**(L-NU/2)
    +4/3*J1*(MB-OM)*R*#E**(-NU)+4*J1*(MB-OM)*R*#E**(L-NU)+4/3*J1*MBP*R**2*#E**(-NU)
    +U*H2*(1/2*EAP**2*V/R*#E**(2*L+NU/2)*(1+V/R*(1-#E**L))+1/2*EAP**2*#E**(2*L+
  NU/2)*(1-P*#E**L*V**2)+EAP*V**2/R**4*#E**(L+NU/2)*(1-#E**L)**2+P*EAP*#E**(2*L+
  NU/2)*(V/R-5/2)-P*EAP*#E**(2*L+NU/2)*V**2*R**(-2)*(1-#E**L)-3/2*EAP/R**2*#E**(L+
  NU/2)*(1+(7/3)*#E**L)+GA*P*#E**(2*L+NU/2)*(3*EAP*V/R+E)+6*GA*P*V/R**3*#E**(L+
  NU/2)*(1-#E**L)+GA*P/R**2*#E**(L+NU/2)*(1-#E**L)+1/2*GES**(-1)*EAP**2*#E**(L+NU/2)*
  ((1-#E**L)/P/R**2-#E**L))
    +U*MB**2*(1/6*EAP**2*R*V*#E**(2*L-NU/2)*(1+V/R)-1/6*EAP**2*V**2*#E**(3*L-NU/2)*
  (1+P*R**2)+1/2*EAP**2*R**2*#E**(2*L-NU/2)*(1+2/3*P*R**2)+1/6*EAP*V**2/R**2*#E**(L-
  NU/2)*(1-#E**L)**2+1/6*EAP*P*V**2*#E**(2*L-NU/2)*(#E**L*P*R**2-2+2*#E**L)-EAP*P*
  R**2*#E**(L-NU/2)*(1+1/2*#E**L)+11/3*EAP*#E**(-NU/2)*(1-3/22*#E**(2*L)+8/11*#E**L)-
  1/3*GA*P*(E**2-P**2)*R**4*#E**(2*L-NU/2)+GA*P*E*R*V*#E**(2*L-NU/2)-8/3*GA*P*
  EAP*R**2*#E**(L-NU/2)*(1-1/4*#E**L)+GA*P*V/R*#E**(L-NU/2)*(1-#E**L)+1/3*GA*P*
  #E**(L-NU/2)+1/3/GES/P*(EAP)**2*#E**(-NU/2)*(1-P*R**2*#E**L)+1/6*GA/GES*EAP**2*R**2*
  #E**(L-NU/2)*(#E**L*(1+P*R**2)-1)-1/6/GES/P*EAP**2*#E**(L-NU/2)*(1+#E**L*(1+P* R**2)))
    +1/12*U*MBP**2*(-EAP*V*R*#E**(L-NU/2)+EAP*V**2*#E**(L-NU/2)*(#E**L-1)+EAP*P*
  R**2*V**2*#E**(2*L-NU/2)+EAP*R**2*#E**(L-NU/2)*(1+2*P*R**2)-GA*P*(2*E+P)*R**4*
  #E**(L-NU/2)-6*GA*P*R*V*#E**(L-NU/2)+GA*P*R**2*#E**(-NU/2)*(#E**L+3))
    +2*U*V2/R**2*#E**(2*L+NU/2)*(((-1/R-V/R**2+V/R**2*#E**L+P*#E**L*V)*EAP-6*GA*
  P/R)*V-EAP)
    +UP*H2*(-GA*P*EAP*V*#E**(2*L+NU/2)+GA/R*(E-P)*#E**(L+NU/2)+2*GA*P*V/R**2*
  #E**(L+NU/2)*(#E**L-1))
    +1/3*UP*MB**2*(2*EAP*R*#E**(-NU/2)+2*GA**2*P**2*R/EAP*#E**(L-NU/2)-GA*P*EAP*R**2*
  V*#E**(2*L-NU/2)+GA*EAP*R*#E**(L-NU/2)*(1+P*R**2)+GA*P*V*#E**(L-NU/2)*((1+P*
  R**2)*#E**L-1)-GA*P*R*#E**(-NU/2)*(2+#E**L))
    +1/6*UP*MBP**2*GA*P*V*R**2*#E**(-NU/2)*(#E**L+NUP*R)
    +4*UP*V2*GA*P*V/R**2*#E**(2*L+NU/2)
\end{verbatim}
\vskip .3in
\begin{verbatim}
DW= -J1*(OM-MB)*R**2*#E**(L/2)*(4*S**(-2)-(2/3)*R**2*#E**(-NU))
    +6*U*H2*#E**(3*(L+NU)/2)*S**(-2)*((1/2)*EAP*(2*V*R**(-2)+(1/3)*S**2*R*#E**(-NU))+GA*
  P*EAP**(-1)*(-P*R**(-1)-(6*V*R**(-1)+1)*(1-#E**(-L))*R**(-3)+EAP*(3*V*R**(-1)+1)*
  R**(-1))+(-2*V*R**(-4)-R**(-3))*(1-#E**(-L))-P*R**(-1))
    +U*MB**2*#E**((3*L+NU)/2)*S**(-2)*(EAP*(2*V+P*R**3+R*(1-#E**(-L))-2*GA*P*R**3+
  (GA/GES)*R*(1-#E**(-L)+P*R**2))-4*GA*P*EAP**(-1)*R**(-1)*#E**(-L)+2*GA*P*(3*V+2*P*
  R**3+2*R-8*R*#E**(-L))-4/R*#E**(-L)+EAP*S**2*R**3*#E**(-NU)*((1/2)*P*R**2-(1/3)*EAP*
  R**2-(11/6)*#E**(-L)+1/2+(1/6)*EAP*(GES*P)**(-1)*(1-#E**(-L)+P*R**2)))
    +U*MBP**2*#E**((L+NU)/2)*S**(-2)*(GA*P/EAP*((1/2)*P*R**3-3*V+(1/2)*R*(1+3*
  #E**(-L)))-GA*P*R**3+(1/2)*P*R**3-V+(1/2)*R*(1-#E**(-L))-(1/6)*EAP*S**2*R**5*
  #E**(-NU))
    -(4/3)*U*MBP*MB*#E**((NU+L)/2)*S**(-2)*R**2*(#E**(-NU)*EAP*R**2*S**2+6*GA*P)
    -24*U*V2*#E**(3*(NU+L)/2)*S**(-2)*R**(-4)*(1+3*GA*P*EAP**(-1))*V
    +6*UP*H2*#E**((L+3*NU)/2)*S**(-2)*R**(-2)*(GA-1-GA*P/EAP)
    +UP*MB**2*#E**((L+NU)/2)*S**(-2)*(2*(GA-1+GA*P*EAP**(-1))+4*(GA*P*EAP**(-1))**2-
  (1/3)*(EAP-GA*P)*R**4*#E**(-NU)*S**2)
\end{verbatim}
\vskip .3in
\begin{verbatim}
DY= +(1/2)*ET*H2*EAP*B*#E**(NU/2)*((1/2)*EAP*V*(1-V*P*R*B+V/R*(1-B))+V*P*(1-
  V/R*(1-B))+C*(V**2/R**3*(B-1)**2-1/R))+(1/12)*ET*MB**2*EAP*R*B*#E**(-NU/2)*(EAP*V*
  (R*(1-V*P*R*B)+V*(1-B))+2*P*V**2*(B-1+(1/2)*P*R**2*B)+C*(V**2*R**(-2)*(B-
  1)**2-1))+(1/24)*ET*MBP**2*EAP*R**2*#E**(-NU/2)*V*(V/R*(B-1+B*P*R**2)-1)+ET*V2*
  EAP*B*#E**(NU/2)*V/R*(V/R**2*(B-1)+B*V*P-1/R)
    +(1/3)*J1*EAP*R*#E**(-NU/2)*(MB-OM)*(1+P*R**2*B+6/R**2/S**2*#E**NU*(B-1+B*R**2*P)-
  5*B)
    +U*H2*B*#E**NU*((1/4)*EAP**3*B*(-3*V/R-V**2/R**2*(1-B)+V**2*P*B-1)+EAP**2*
  (V/R**3*(2+B)-(1/2)*V**2/R**4*(3-4*B+B**2)+(P*R**2)**(-1)*(-S**2*#E**(-NU)+(1/4)*
  R**(-2)*(6+B-7*C))+(1/2)*P*B*(2-V/R+V**2/R**2*(3-B))+(1/4)/R**2*(5+7*B)+3*
  #E**NU*S**(-2)*R**(-4))+EAP*(8*V*R**(-5)*(C-1)-2*V**2*R**(-6)*C*(B-1)**2-P*B*(3*
  (R*S)**(-2)*#E**NU*P)+P*(-2*V*R**(-3)-(5/4)*P*B+2*V**2*R**(-4)*(1-B)-(1/2)*
  R**(-2)*(1+5*B)+6*#E**NU/S**2/R**4*(1-B))+5*S**2*#E**(-NU)/R**2-(1/2)*R**(-4)*(15+
  (5/2)*B-(19/2)*C)+3*#E**NU/S**2/R**6*(C-B)))
    +U*H2*B*#E**NU*((1/2)*GA*EAP**2*R**(-3)*(-V*(1-B+7*P*R**2*B)+R*(1+B-2*P*R**2*
  B))+GA*EAP*(-V*C*R**(-5)*(B-1)**2+P**2*V/R*B*(1+3*V/R)+P**2*B*(2+3*
  #E**NU/R**2/S**2)+P*(V/R**3*(1+8*B)-3*V**2/R**4*(1-B)+2/R**2*(1+(3/4)*B)+3*
  #E**NU/S**2/R**4*(2+B))-R**(-4)*(1+(1/2)*B-(3/2)*C))+GA*P**2*(-2*V*R**(-3)*(2+B)+
  6*V**2*R**(-4)*(1-B)-2*B/R**2-P*B*(1+3*#E**NU/R**2/S**2)+6*#E**NU/S**2/R**4*(1-B))+
  GA*P*(-6*V**2*R**(-6)*C*(B-1)**2-2*V*R**(-5)*(10+B-11*C)+R**(-4)*(3*C-B-14)+3*
  #E**NU/S**2/R**6*(5*C-4-B))+GAP*P*(-3*V/R**2*EAP-(EAP-P)/R+(6*V+R)/R**4*(1-C)))
    +U*H2*B*#E**NU*(+EAP**2/(GES*P*R**4)*(S**2*R**2*#E**(-NU)-3/2-(1/4)*B+(7/4)*C)-
  (1/2)*EAP**2/GES/R**2*(1+B+B*P*R**2/2)+(1/4)*EAP**3/GES*((1+B)/P/R**2+B)+(1/2)*GA*
  EAP**2/GES/R**2*(2+B*(1+P*R**2)))
    +U*MB**2*(EAP**3*R**2*B*(-(1/4)*V/R*B-(1/12)*(V/R)**2*B*(1-B*(1+P*R**2))-(1/12)*
  (1+B*(2+P*R**2)))+EAP**2*B*((5/6)*V/R*(1+B/5)+(1/2)*(V/R)**2*(-5/6+B-B**2/6)+(P*
  R**2)**(-1)*(-7*C/12+B/12+1/2)-(1/3)*S**2/P*#E**(-NU)*(1+2*P*R**2)-(1/12)*P**2*R**4*
  B+P*R**2*(1+B/6)+(1/6)*P*V*R*B+(1/2)*P*B*V**2*(1-(1/3)*B*(1+P*R**2/2))-P/S**2*
  #E**NU*(1-B*(1+P*R**2/2))-C/4+2+B/3+(1/2)/S**2/R**2*#E**NU*(C+B))+EAP/R**2*B*((4/3)*
  V/R*(C-1)-2/S**2*#E**NU*P+2/3*P*V**2*(1-B)-1/3*C*V**2/R**2*(B-1)**2+P*R**2*(-
  7/6-B*P*R**2/4-B/2-4/3*V/R-1/3*B*P*R**2*(V/R)**2)+89*C/12-13/6-B/4-2/S**2/R**2*
  #E**NU*(1-C)+S**2*R**2*#E**(-NU)))
    +U*MB**2*(+GA*EAP**2*B*(1/6*V/R*(B-1)+1/6*EAP*R**2*(1-B+2*B*P*R**2)+1/3*(B+4*
  C-4)+P*R**2*(10/3*(1-B/4)-7/6*(P*R**2+V/R)*B+#E**NU/S**2/R**2*(1-B*(1+P*R**2))))+
  GA*EAP*B/R**2*(-1/6*V/R*C*(B-1)**2+(P*R**2)**2*(3/2*B*V/R*(1+2/3*V/R)-14/3+7/3*B
  -10/S**2/R**2*#E**NU*(1-2/5*B))+2*(P*R**2)**3*B*(7/12+#E**NU/S**2/R**2)+P*R**2*(5/3*
  V/R*(1+4/5*B)+(V/R)**2*(B-1)+19/6*C-1+7/6*B-#E**NU/S**2/R**2*(7-2*B-8*C))+1/6* (C-1)))
    +U*MB**2*(+GA*P**2*B*(-10/3*V/R*(1+B/5)+2*(1-B)*(V/R)**2+1/3*(2+B))-2*
  P/S**2/R**2*#E**NU*B*GA*P+1/6*GA*P**3*R**2*B**2*(1-2*V/R*(1+3*V/R))+GA*P/R**2*
  (1/3*V/R*(11-B**2-10*B)-(V/R)**2*(B-1)**2+17/6+B**2/6+4*B/3+2/S**2/R**2*#E**NU*(1-
  B))-S**2*B*#E**(-NU)*2*GA/3*P+GAP*P/R*(1/3*((EAP-P)**2-P**2)*R**4*B-(EAP-P)*B*V*
  R+8/3*EAP*R**2*(1-B/4)+V/R*(B-1)-1/3))
    +U*MB**2*(+EAP**2/GES*(+(1/2)*(P*R**2)**(-1)*(7/6-B**2/6-B)+1/3*S**2/P*B*#E**(-NU)-
  1/6*B*(1+B*(1+P*R**2/2)))+1/12*EAP**3/(GES*P)*(1+B-P*R**2*B**2*(1+P*R**2))+GA*
  EAP**2/GES*(1/2*(P*R**2)**2*B**2*(4/3+#E**NU/S**2/R**2)-8/3*P*R**2*B*(1-B/2)-P/S**2*
  B*#E**NU*(1-B)+5/3+2/3*B**2-11/6*B+1/2/S**2/R**2*#E**NU*(B-1)**2)+1/3*GA*
  EAP**3/GES*R**2*B*(1-3/2*B*(1+P*R**2))+1/6*GAP/GES*EAP**2*R*(1-B*(1+P*R**2))+1/6*
  GA*EAP**3/GES**2/P*((B-1)**2-2*P*R**2*B*(1-B*(1+P*R**2/2)))-1/6*GA*EAP**2*
  GESP/GES**2*R*(1-B*(1+P*R**2)))
    +U*MBP**2*((1/24)*EAP**2*R**2*(-1+V**2/R**2*B*(1-B)+3*V/R*B-P*B*(V**2*B+R**2)+
  2*GA/GES*(1-B*(1+P*R**2)))+EAP*(1-V/R/2+(B-7*C)/12+(1/6)*V**2/R**2*(1-B)-
  (1/2)/S**2*#E**NU*P*(1-B*(1+P*R**2/2))+(1/4)/S**2/R**2*#E**NU*(C+B-4)+(1/12)*(P*
  R**2)**2*B+(1/6)*P*R**2*(1+B*(1-(V/R)**2))-(1/3)*S**2*R**2*#E**(-NU))+(1/12)*GA*
  EAP**2*R**2*(1-B*(1-2*P*R**2))+(1/12)*GA*EAP*(-1-3*C/2+B/2+V/R*(1-B)+
  6/S**2/R**2*#E**NU*P*R**2*(1-B*(1+P*R**2))-5*(P*R**2)**2*B/2+P*R**2*(7*V/R*B+4-
  2*B))+(1/6)*GA*P**2*R**2*B*(1+P*R**2/2-V/R-3*(V/R)**2)+(1/12)*GA*P*(26+B+19*C-
  16*V/R*(1+B/8)+6*(V/R)**2*(1-B)-12/S**2/R**2*#E**NU*(1-B/4+3*C/4))+(1/2)*GA*
  P**2/S**2*#E**NU*(1+B*(1+P*R**2/2))-(1/12)*GAP*P*R*(1+3*C-6*V/R-(2*E+P)*R**2))
    +U*MB*MBP*(EAP**2*R*((2/3)*(1+B*(1+P*R**2))+(2/3)*GA*(1-B*(1-3*P*R**2))+GA/GES*
  (1-B*(1+P*R**2)))+EAP*(4/R+GA*P*R*(7/3+4/S**2/R**2*#E**NU)*(1-B*(1+P*R**2))+
  (4/3)*GAP*P*R**2)+(4/3)*GA*P/R)
    +U*V2*B*#E**NU*(EAP**2*B/R**2*(1+3*V/R+V**2/R**2*(1-B*(1+P*R**2)))+EAP/R**4*(-12*
  V/R+4*V**2/R**2*(1-B*(1+P*R**2))+4*(1-GA)-12*#E**NU/S**2/R**2+2*GA*V/R*(1-B*(1-
  7*P*R**2)))+GA*P*R**(-4)*(12-4*V/R*(8+B*(1+P*R**2))+12*V**2/R**2*(1-B*(1+P*
  R**2))-36*#E**NU/S**2/R**2)+12*GAP*P*V/R**4)
    +UP*H2*#E**NU*(EAP*R**(-3)*(3/S**2/R**2*#E**NU*((1-GA)*(1-B)-B*P*R**2)-2*GA*V/R*
  (1-B)-GA*P*R**2*B*(1+V/R*(1+B))-(3/2)*GA*(1-B))+(1/2)*GA*EAP**2*(-2*B*V/R**2-
  B/R+P*V*B**2+(1-B)/P/R**3*(1-1/GES))+3*GA*P*(EAP-P)*B*#E**NU/S**2/R**3-GA*P*(1-
  B)/R**3*(-3*#E**NU/S**2/R**2+4*V/R))
    +UP*MB**2*(EAP**2*((1/6)*R*(1+B*(1+P*R**2))-(1/3)*GA*V*B*(1-(1/2)*B*P*R**2)+
  (1/6)*GA/(P*R)*(1-B)*(1-1/GES)+(1/3)*GA*R*(1/2-B*(1-P*R**2))+(1/6)*GA*R/GES*(1-
  B*(1+P*R**2)))+EAP*((1-GA)/S**2/R**3*#E**NU*(1-B*(1+P*R**2))-(1/3)*GA*V/R**2*(1+
  B*P*R**2/2-B*(1-B*P*R**2/2)+(P*R**2*B)**2/2)+GA*P*R*(7/3-B*(1/2+2*P*R**2/3))-
  (3/2)*GA/R*(1-B/3))+GA*P**2*B*(2*V/3+#E**NU/S**2/R)-(1/3)*GA**2*P/R*(1-B*(1+P*
  R**2))*(1-1/GES)-(2/3)*GA*P/R*(V/R*(1-B)+1+(3/2)/S**2/R**2*#E**NU*(1-B))+EAP**(-1)*
  (-(2/3)*GA*GAP*P**2-(5/3)*(GA*P)**2/R*(1-B/5)-2*(GA*P)**2*#E**NU/R**3/S**2*(1-B)+
  (1/3)*GA**2*P**3*R*B*(1+6*#E**NU/R**2/S**2)))
    +UP*MBP**2*((1/6)*GA*EAP*(V*(1-B*P*R**2/2)-R*(1-2*P*R**2)/2+R*C/2)+(1/3)*GA*P*
  (V-R*C))
    -(2/3)*UP*MB*MBP*(GA*EAP*(1-2*P*R**2)+2*(GA*P)**2/EAP)
    +2*UP*V2*GA*V/R**4*B*#E**NU*(EAP*(2-B*P*R**2)+4*P)
\end{verbatim}
\eject
\end{document}